\documentclass{article}

\usepackage{epsfig}
\usepackage{epsf}
\def\e{\begin{equation}}
\def\f{\end{equation}}
\def\=#1{\overline{\overline{#1}}}
\def\_#1{{\bf #1}}
\def\-#1{{\bf #1}}
\def\o{\omega}

\def\.{\cdot}

\def\l#1{\label{eq:#1}}
\def\r#1{(\ref{eq:#1})}

\def\l#1{\label{eq:#1}}
\def\r#1{(\ref{eq:#1})}

\title{Frequency range and explicit expressions for negative permittivity and permeability for an isotropic  medium formed by a lattice of perfectly conducting $\Omega$ particles
}
\vspace{2in}
\author{Constantin R. Simovski$^1$,
and Sailing He$^{2,3}$  \\
$\mbox{ }^1$ Department of Physics, \\
Institute of Fine mechanics and Optics, 197101 St. Petersburg, Russia;\\
$\mbox{ }^{2}$Centre for Optical and Electromagnetic Research,\\
State Key Laboratory of Modern Optical Instrumentation, \\
Zhejiang University, Yu-Quan, 310027 Hangzhou, P. R. China;\\
$\mbox{ }^3$ Department of Electromagnetic Theory, \\
Royal Institute of Technology,  S-100 44, Stockholm, Sweden.}

\begin{document}
\baselineskip 6mm
\maketitle

\newpage
\section*{\bf Abstract}

An analytical model is presented for a rectangular lattice of
isotropic scatterers with electric and magnetic resonances. Each
isotropic scatterer is formed by  putting appropriately 6
$\Omega$-shaped perfectly conducting particles on the faces of a
cubic unit cell. A self-consistent dispersion equation is derived
and then used to calculate correctly the effective permittivity
and permeability in the frequency band where the lattice can be
homogenized.  The frequency range in which both the effective
permittivity and permeability are negative corresponds to the
mini-band of backward waves within the resonant band of the
individual isotropic scatterer.

PACS: 41.20.Jb, 78.20.Ci, 42.25.Bs, 42.70.Qs.

\newpage

\section{Introduction}

In 1967 Veselago considered an isotropic continuous medium with
both negative permittivity and negative permeability and proved
that a negative refraction would occur   when a monochromatic wave
impinges on the surface of such a medium  \cite{Veselago}. As a
result he obtained theoretically a quasi-lens (the lens does not
focus parallel rays) from a slab of such a hypothetic material. In
the Veselago theory this quasi-lens must form an image of any
source without distortion (since each interface is a parallel
plate). The phase of the radiated field at the focal point should
be the same as the phase of the source dipole. These results are
the consequences of backward waves in such a medium. The phase
velocity and the Pointing vector are in opposite direction for a
backward wave. In 2000  Pendry claimed that the Veselago
quasi-lens must possess an extraordinary property to amplify and
focus the evanescent waves in the spatial spectrum of a point
source and therefore this lens must be perfect \cite{lens}. Since
then much attention has been attracted to this research area both
theoretically and experimentally (see e.g. \cite{Science}$-$\cite{negative}).
Such negative media  do not exist in nature but
have already got different names in the literature such as
backward-wave media, the Veselago media, left-handed media and
negative meta-materials. Since it is not clear yet how to make
such media in the optical range of frequencies, the experimental
efforts have been concentrated on creating negative meta-materials
at microwave frequencies. A structure allowing the propagation of
backward  waves within a certain frequency band has been suggested
in \cite{Smith} and then experimentally studied in \cite{Science},
where the negative refraction phenomenon was primarily observed.
However, the structure of \cite{Smith} (a combination of two
lattices: a lattice of infinitely long parallel wires and a
lattice of the so-called {\it split-ring resonators}) is not
isotropic.   This structure can be described with negative
permittivity and permeability only if the propagation direction is
orthogonal to the axes of wires \cite{book}.

To the best of our knowledge, the only isotropic medium with
possible negative permittivity and permeability (within a certain
frequency band) that has been reported is formed by many isotropic
cubic cells of $\Omega$ particles \cite{Bruno}. Each isotropic
cubic cell  is made by putting 6 $\Omega$-shaped perfectly
conducting particles on its faces (as shown in Fig. 1). Fig. 1
shows how  to make an isotropic unit cell by putting successively
3 pairs of $\Omega$-particles on the opposite faces of a cubic
unit cell in an appropriate way. Then each cubic cell can be
described approximately as an isotropic resonant scatterer. The
frequencies at which both the effective permittivity and permeability
of such a composite medium are  negative are within  the
resonant band of each individual isotropic  scatterer. Note that
conducting $\Omega$-particles were first suggested in \cite{JE}
for creating the bianisotropic composites.

In the present paper we consider a negative meta-material formed
by a rectangular lattice of isotropic cubic unit cells of $\Omega$
particles (as shown in Fig. 1). We introduce an analytical model
to describe the dispersion properties of the lattice and calculate
correctly the effective permittivity $\epsilon_{eff} (\o )$ and
permeability $\mu_{eff} (\o )$ in the frequency band where the
lattice can be homogenized. We compare the results obtained by the
present model with the previous results obtained by  the Maxwell
Garnett model, and show that the latter can not give reliable
values for either the real parts or the  imaginary parts of the
effective material parameters $\mu_{eff}$ and $\epsilon_{eff}$.

From the negative dispersion of the lattice at low frequencies we
obtain the frequency range of backward waves within the resonant
band of the isotropic scatterers. In this frequency range the
phase velocity is in an opposite direction of the group velocity.
Our calculations show that this backward wave range is exactly the
same as the frequency range within which both $\epsilon_{eff}$ and
$\mu_{eff}$ are negative.

The case when the lattice is formed by many parallel bianisotropic
$\Omega$-particles is also considered in an appendix. It is shown
that the bianisotropy of the lattice particles  reduces the
possibility for the existence of backward waves.

\section{Polarizability of an isotropic cubic unit cell}

For bianisotropic particles (a $\Omega$-particle is a special
case), one has the following four dyadic polarizability matrices
relating the electric dipole moment $\-p$ and the magnetic dipole
moment $\-m$ with the local electric and magnetic
fields:\begin{eqnarray}
 \_p &=&\=a_{ee}\. \_E^{\rm loc}   +\=a_{em}\.  \_H^{\rm
loc}
\l{p} , \\
\_m & =  & \=a_{me}\.  \_E^{\rm loc} +  \=a_{mm}\. \_H^{\rm loc}
\l{m} .
\end{eqnarray}
For reciprocal particles, one has $\=a_{me}=-\=a_{me}^T$.

For a cubic unit cell of $\Omega$ particles as shown in Fig. 1,
the situation is quite different. The bianisotropy cancels out
since e.g. each pair of opposite $\Omega$ particles shown in
Fig. 1 have equal absolute value but opposite sign for $\=a_{em}$.
The polarizabilities $\=a_{ee}$ and $\=a_{mm}$ of a cubic unit
cell of 6 $\Omega$-particles (see Fig. 1) are the sums of the
polarizabilities $\=a_{ee}^{\Omega}$ and $\=a_{mm}^{\Omega} $ for
individual particles \cite{Bruno}. The influence of the mutual
coupling of the particles is small and the resonant frequency for
such a cubic unit cell of 6 $\Omega$-particles has only a small
shift from that for an isolated $\Omega$-particle. The resonant
excitation by an electric field directed along the arms of
$\Omega$ particles is mainly due to the presence of the arms. The
resonant behavior  retains for the total electric polarization of
the two opposite omega particles.  Within the resonant band, the
quasi-static polarizability component $a_{ee}^{zz\Omega}$ (which
describes the response of the  particle to an electric field
normal to the direction of the arms; see Fig.  1(a)) of the
$\Omega$ particle is small as compared to $a_{ee}^{yy\Omega}$. Such a
cubic unit cell can be therefore considered as an isotropic
scatterer and thus one has
$a_{ee}^{xx}=a_{ee}^{yy}=a_{ee}^{zz}=a_{ee}$ and
$a_{mm}^{xx}=a_{mm}^{yy}=a_{mm}^{zz}=a_{mm}$ \cite{Bruno}. One can
easily show that \cite{Bruno} \e a_{ee}\approx 2a_{ee}^{xx\Omega},
\qquad a_{mm}=2a_{mm}^{yy\Omega} \l{omegas}.\f The expressions for
the parameters $a_{ee}^{xx\Omega}$ and $a_{mm}^{yy\Omega}$ have
been given in \cite{JEWA}.  In the present paper we consider a material formed by a rectangular
lattice of such scatterers (see Fig. 1(b)).

The electric and magnetic polarizabilities for a cubic cell of
$\Omega$-particles are calculated (cf. \cite{Bruno}) and shown in
Fig. 2 as a numerical example. The geometric parameters for the
$\Omega$-particles are chosen as $r=1.5$ mm, $w=0.4$ mm, $h=0.2$
mm and  $l=2$ mm (see Fig. 1(a)). The size of the cubic unit  cell
is $4$ mm. The relative permittivity for the background medium is
chosen as  $\epsilon_b=1.5$ (the permeability of the background
medium is assumed be the same as the one for vacuum in the present paper).
Resonances of $a_{ee}$ and $a_{mm}$ (frequencies at which the real
parts of these parameters become zero) occur at the frequency of
8.05 GHz and 8.14 GHz, respectively. We have chosen the time
dependence $e^{j\o t}$, and thus the imaginary parts of these
parameters are  negative. Below we consider a material formed by a
rectangular lattice of such isotropic scatterers (see Fig. 1(b)).

\section{Wrong results predicted by the Maxwell Garnett model for the regular lattice}

 In \cite{Bruno},  the well-known Maxwell
Garnett model was used and  the following expressions for the material parameters  were obtained,
\e
\epsilon_{eff}=\epsilon_{b}+{1\over F}\left({Na_{ee}\over \epsilon_{0}}-\
{N^2a_{ee}a_{mm}\over 3\epsilon_{0}\epsilon_{b}\mu_0}
\right),
\l{MG1}\f
\e
\mu_{eff}=1+{1\over F}\left({Na_{mm}\over \mu_{0}}-\
{N^2a_{ee}a_{mm}\over 3\epsilon_{0}\epsilon_{b}\mu_0}
\right),
\l{MG2}\f
where
$$
F=1-{Na_{ee}\over 3\epsilon_{0}}-{Na_{mm}\over 3\mu_{0}}-
{N^2a_{ee}a_{mm}\over 9\epsilon_{0}\epsilon_{b}\mu_0}.
$$

In the Maxwell Garnett model, one only needs to know the scatterer
polarizabilities and the density $N$ of the scatterers. Thus the
obtained result for a regular lattice of  scatterers  will be the
same as that for a random distribution of scatterers if the
density of the scatterers is the same. For the case considered in
the present paper (i.e., a rectangular lattice of isotropic  cubic
unit cells as shown in Fig. 1(b) with periods $d_x,d_y,d_z$ along
the Cartesian axes),   one has $N=1/d_xd_yd_z$. The frequency
dependencies of $\epsilon_{eff}$ and $\mu_{eff}$  predicted by the
Maxwell Garnett model  are shown in Fig. 3 when $d_x=d_y=d_z=d=8$
mm (i.e. $N=(1/8^3)\cdot 10^{-9}$ m$^{-3}$). From this figure one
sees that both ${\rm Re}(\mu_{\rm eff})$ and ${\rm
Re}(\epsilon_{\rm eff})$ are negative within a certain frequency
band (8.34$-$8.62 GHz). However, within this frequency band the
imaginary parts of $\mu_{\rm eff}$ and $\epsilon_{\rm eff}$ have
high values (about $-4\sim -6$). The radiation losses are so large
that the obtained negative effective permittivity $\epsilon_{eff}$
and permeability $\mu_{eff}$  have no practical importance for
applications.

The above results predicted by the Maxwell Garnett model (in
\cite{Bruno} for a random arrangement of cubic unit cells)  are
not correct for the present case of the regular lattice. Within
the resonant band (7.6-8.6 GHz) of each $\Omega$-particle,  the
wavelength ($\lambda=28-33$ mm) in the background medium
($\epsilon_b=1.5$) is much larger than the distance $d$ (8 mm)
between the isotropic cubic scatterers, and this distance is
larger than the size (4 mm) of each  scatterer. From this point of
view one may expect that the homogenization is possible. However,
the results predicted by the Maxwell Garnett model are
inconsistent near the resonance of the scatterer. The Maxwell
Garnett model  is based on the Clausius-Mossotti relations for the
electric and magnetic fields. These relations for cubic lattices
of static dipoles have been found as good approximations in the
works of Sivukhin \cite{Sivukhin} (for an infinite lattice and for
a half-space) and McPhedran \cite{McPhedran} (for arrays of finite
sizes). The Clausius-Mossotti relations were also confirmed to be
accurate enough for cubic lattices of dipoles in a time-harmonic
case \cite{We} under the condition that the wavelength is large
compared to the lattice period $d$. However, the resonant case is
not considered in these works.   In the resonant  case, the
Maxwell Garnett model may give very high absolute values for the
real parts of $\epsilon_{eff}$ and $\mu_{eff}$ within the resonant
band, and this would lead to a dramatic shortening of the
wavelength in the homogenized medium. The wavelength in the
homogenized medium becomes comparable with the particle distance
$d$ and thus the Maxwell Garnett model becomes contradictory at
some frequencies.

In the present case (see Fig. 3) the absolute values of real parts
of  $\epsilon_{eff}$ and $\mu_{eff}$ are so high within the narrow
band 8.37-8.40 GHz where the Maxwell Garnett model becomes
contradictory. Outside this narrow band $|{\rm
Re}(\epsilon_{eff})|$ and $|{\rm Re}(\mu_{eff})|$ are not very
high. However, the inconsistency of the Maxwell Garnett  model for
the present case of regular lattice is evident over the whole
resonant band of $\Omega$-particles (8.1-8.5 GHz). In Fig. 3 one
can see that the imaginary parts of $\epsilon_{eff}$ and
$\mu_{eff}$ are rather significant at those frequencies. On the
other hand,  we know that ${\rm Im}(\mu_{\rm eff})$ and ${\rm
Im}(\epsilon_{\rm eff})$ must be identically zeros physically if
the lossless scatterers are arranged in a regular lattice. The
cancelation of the radiation resistances of the scatterers (due
to their electromagnetic interaction) is obvious for an infinite
regular array and thus there is no radiation loss at all (see e.g.
\cite{Sipe}).

In the present paper, we introduce another homogenization model
which gives correctly zero value for the imaginary parts of
$\epsilon_{eff}$ and $\mu_{eff}$.  A comparison in Section 5 will
show that the real parts of $\epsilon_{eff}$ and $\mu_{eff}$
predicted by the Maxwell Garnett model are also wrong within the
resonant band of the scatterers.

\section{Dispersion equation for the lattice of the isotropic scatterers}

Consider the rectangular lattice of the isotropic scatterers (shown in Fig. 1)
with periods $d_x,d_y$ along the $x$ and $y$ axes and $d_z=d$ along the $z$ axis. In the present case,
the electric dipole moment  is  orthogonal to the magnetic dipole moment for each scatterer.   Let $\-p$ and $\-m$ denote the dipole
moments of a reference scatterer located at the origin.
 Thus one has
\begin{eqnarray}
\_p &=& a_{ee} \_E^{\rm loc},
\l{p1} \\
\_m & =  & a_{mm} \_H^{\rm loc}.
\l{m1}
\end{eqnarray}
We need to find the eigenmodes for such a lattice of isotropic
scatterers. Assume  the wave propagates along the $z$-axis. Let
$\beta$ denote the propagation constant of an eigenmode. The first
Brillouin zone is the interval $0\le \beta d\le \pi$. Then
$\-p(n_x,n_y,n_z)$ and $\-m(n_x,n_y,n_z)$ (the electric and
magnetic dipole moments of a scatterer located at a lattice node
with coordinates $x=n_xd_x,y=n_yd_y,z=n_zd$) can be expressed
through $\-p$ and $\-m$ (for the reference scatterer) as
\begin{eqnarray}
\_p(n_x,n_y,n_z) &=& \_pe^{-jn_z\beta d},
\l{p2} \\
\_m(n_x,n_y,n_z) & =& \_me^{-jn_z\beta d}.
\l{m2}
\end{eqnarray}

Since eigenwaves are linearly polarized in  the present case,   we
can assume without loss of generality that $\_p=p\-x_0$ and
$\_m=m\-y_0$ (see also Fig. 1). Then the local electric and
magnetic fields are directed along the  $x$  and $y$ directions,
respectively (the local field has the same polarization as
the eigenmode).
This  allows the following scalar expressions for
the local field amplitudes in terms of  $p$ and $m$:
\begin{eqnarray}
E^{\rm loc} &=& Ap+Bm,
\l{ploc1} \\
H^{\rm loc} &=& Cm+Dp,
\l{mloc1}
\end{eqnarray}
where $A,B,C,D$ are the so-called interaction factors of the
lattice,  which depend only on the lattice geometry, the frequency
$\o$ and the propagation factor $\beta$ (they do not depend on the
polarizabilities of the scatterers). From the reciprocity it
follows that $B=D$. From the duality it follows that $C={A /
\eta^2}$, where $\eta= \sqrt{\mu_0/\epsilon_b\epsilon_0}$ is the
wave impedance of the background medium.

From Eqs. \r{p1},\r{m1}, \r{ploc1} and \r{mloc1}, one obtains  the
following  two important relations, \begin{eqnarray} p(1-a_{ee}A)
&=& a_{ee}Dm,
\l{p3} \\
m(1-{a_{mm}A\over \eta^2}) &=& a_{mm}Dp.
\l{m3}
\end{eqnarray}
Denote the dimensionless ratio
$\eta p/m$ as $\alpha$, then we have
 \e \alpha\equiv \eta{ p\over
m}={\left({\eta\over a_{mm}}-{A\over \eta}\right)\over D} .
\l{alp}\f This formula will be used in the next section to find
the effective material parameters of the lattice at low
frequencies.

From Eqs. \r{p3} and \r{m3} one  obtains \e \left({1\over
a_{ee}}-A\right)\left({\eta^2\over a_{mm}}-A\right)=\eta^2D .
\l{disp}\f The relation \e {\rm Im}\left({1\over
a_{ee}}\right)={\o^3(\epsilon_0\mu_0)^{3\over 2}\epsilon_b^{1\over
2}\over 6 \pi \epsilon_0} \l{sip}\f was first obtained in
\cite{Sipe} for a particle re-radiating the light and later was
reproduced as the consequence of the energy conservation for a
lossless dipole scatterer (see e.g. \cite{JOSA}). For lossless
magnetic scatterers, one has the following corresponding form,
$$
{\rm Im}\left({1\over
a_{mm}}\right)={\o^3(\epsilon_0\epsilon_b\mu_0)^{3\over 2}\over 6
\pi \mu_0}.
$$
For the interaction factor $A$, an explicit approximation has been
found in \cite{JOSA}: \e A={\eta\o\over 2 d_xd_y}\left(q_0+{\sin
kd\over \cos kd - cos \beta d} \right)+j{k^3 \over 6 \pi
\epsilon_0\epsilon_b}, \l{exact}\f where
$k=\o\sqrt{\epsilon_0\mu_0\epsilon_b}$ is the wavenumber in the
background medium, and $q_0$ is the real part of the dimensionless
interaction factor of a 2D grid of dipoles with periods $d_x,d_y$.
$q_0$ has the following explicit  expression when $d_x=d_y=a$
\cite{JOSA},
\e q_0={1\over 2}\left({\cos kas\over kas}-\sin
kas\right), \l{q0}\f
where $s \approx 1/1.4380=0.6954$. Relation
\r{exact} is almost exact  when $d\gg d_x,d_y$, and  has only a
small error when $d=d_x=d_y$ \cite{JOSA}.

Since
$$
{\rm Im}A={\rm Im}\left({1\over a_{ee}}\right)={\rm Im}\left({\eta^2\over a_{mm}}\right),
$$
the left-hand side of \r{disp} is real.  Thus we have \e {\rm
Re}\left({1\over a_{ee}}-A\right){\rm Re}\left({\eta^2\over
a_{mm}}-A\right)=\eta^2D \l{real}.\f Below we show that $D$ is
also real and we derive its explicit expression. From the
definition of  $D$, one has \e H_y^{\rm
loc}(x=0,y=0,z=0)=\sum\limits_{n_x}\sum\limits_{n_y}\sum\limits_{n_z}
H_y(n_xd_x,n_yd_y,n_zd)\equiv Dp , \l{def}\f
where
$H_y(n_xd_x,n_yd_y,n_zd)$ is the $y$-component of the magnetic
field at the origin produced by the $x$-polarized electric dipole
with dipole moment $\-p(n_x,n_y,n_z)=p(n_x,n_y,n_z)\-x_0$. From
the reciprocity we know that $H_y(n_xd_x,n_yd_y,n_zd)$ is equal to
the field at the point $(x=n_xd_x,y=n_yd_y,z=n_zd)$ produced by
the reference dipole $\-p$. In Eq. \r{def} the summation is over all
integers $(n_x,n_y,n_z)$ from $-\infty$ to $\infty$ except
$n_x=n_y=n_z=0$. Here we  use the following plane-wave
representation for the magnetic field produced by an electric
dipole $\-p=p\-x_0$ located at the origin \cite{Clemmow}: \e
H_y(x,y,z)={j\o\over
8\pi^2}\Psi(z)p\int\limits_{-\infty}^{\infty}\int\limits_{-\infty}^{\infty}
e^{-j(q_xx+q_yy)-\Psi(z)\sqrt{k^2-q_x^2-q_y^2}z}dq_xdq_y,
\l{Clem}\f where $\Psi(z)=+1$ for $z>0$, $\Psi(z)=-1$ for $z<0$
and $\Psi(z)=0$ for $z=0$. Substituting Eqs. \r{Clem} and  \r{p2} into
definition \r{def} and changing the order of the summation over
$n_z$ and the integration over $q_x$ and $q_y$, one obtains
$$
D={2j\o\over 8\pi^2}\sum\limits_{n_x}\sum\limits_{n_y}\int\limits_{-\infty}^{\infty}\int\limits_{-\infty}^{\infty}
e^{-j(q_xx+q_yy)}\sum\limits_{n_z=1}^{\infty}e^{-jn_z\beta d}e^{-n_zd\sqrt{k^2-q_x^2-q_y^2}}dq_xdq_y .
$$
Summarizing the geometrical series and using known explicit
formulas for the integration  over $q_x$ and $q_y$, one can obtain
the following final result for $D$, \e D=-{\o\over
2d_xd_y}\sum\limits_{n_x}\sum\limits_{n_y}{\sin \beta d\over \cos
\sqrt{k^2-\left({2\pi n_x\over d_x}\right)^2-\left({ 2\pi n_y\over
d_y}\right)^2}d-\cos\beta d} . \l{DDD}\f

Numerical calculations have shown
that the contribution from the terms with $n_x,n_y=\pm 1,\pm 2,\dots$
becomes significant only if $d_x,d_y\gg d$. These terms correspond
to the effect of the evanescent Floquet spatial harmonics
generated by the 2D grids of dipoles with $n_z=\pm 1,\pm 2\dots$ (the
reference dipole is located in the 2D grid with $n_z=0$) . If we
consider the case $d_x=d_y=a\le d$, then we can neglect these
terms and take into account only the plane-wave interaction between
the 2D grids of dipoles. For such a case, one has the following real-valued expression for $D$,
 \e D=-{\o\over
2d_xd_y}{\sin \beta d\over \cos kd-\cos\beta d} .\l{DD}\f Note
that relation \r{exact} was derived in \cite{JOSA} under the same
condition. Formulas \r{exact} and \r{DD} are both approximate but
self-consistent. Therefore, these relations lead to the
real-valued dispersion equation \r{real}.

Substituting \r{exact} and \r{DD} into \r{real},  one obtains $$
\cos^2 \beta d (1+\gamma_1\gamma_2)- \cos \beta d
\left(2\gamma_1\gamma_2\cos kd -(\gamma_1+\gamma_2)\sin kd\right)
$$
\e
- (1-\gamma_1\gamma_2)\cos^2 kd-(\gamma_1+\gamma_2)\sin kd\cos
kd=0, \l{disp1}\f where
$$
\gamma_1={{\rm Re}\left({1\over a_{ee}}\right)2d_xd_y\over \eta
\o}, \qquad
\gamma_2={{\rm Re}\left({\eta^2\over a_{mm}}\right)2d_xd_y\over
\eta \o}.
$$
The dispersion relation \r{disp1} has two roots for a fixed
frequency. One of the roots satisfies the following condition
(obtained from Eqs. \r{p3}and \r{m3})
$$
{\left({\eta\over a_{mm}}-{A\over \eta}\right)\over D}={\eta
D\over \left({1\over a_{ee}}-A\right)}.
$$
This root is $\beta$ that we show in  our dispersion plots below.
The other root does not satisfy the above condition and thus is
spurious.

Fig. 4 shows the dispersion for the same cubic lattice of
isotropic scatterers as used for Fig. 3. From this figure one can
see that there is a band (between 8.14 and 8.37 GHz) in which the
group velocity is negative. This is the backward-wave band. Since
the Poynting vector in a lossless medium must be directed along
the group velocity, this band corresponds to the case when the
energy transports oppositely to the phase.   Of course, this
situation would be trivial for some high-frequency dispersion
branches. For a lattice of dipoles, negative dispersion is
inherent for every even dispersion branch ($\pi (2n+1)<kd<\pi
2(n+1)\pi$, where $n=0,1,2\dots$). However, only in the first
frequency zone (where $0<kd<\pi$) the lattice period is smaller
than $\lambda/2$ (half of the wavelength in the background medium)
and thus the homogenization of the  lattice is possible. Only if
the lattice of scatterers can be homogenized, the material
parameters $\epsilon_{eff}$ and $\mu_{eff}$ can be introduced  and
the lattice of scatterers can be interpreted as  a continuous
medium. It the Veselago theory \cite{Veselago} the backward wave in a
continuous medium should correspond to a case when both
$\epsilon_{eff}$ and $\mu_{eff}$ are negative. The negative
dispersion in the homogenized structure implies the backward wave
and therefore our $\epsilon_{eff}$ and $\mu_{eff}$ should be
negative if the Veselago theory  is applicable for
dispersive media.

Fig. 5 gives an enlarged view of Fig. 4 for the frequency
dependence of ${\rm Re}\beta$ and ${\rm Im}\beta$ in the frequency
range 7-10 GHz. From this figure one sees that there are two
stopbands within the resonant band of each scatterer (cf. Fig. 2).
The lower one (stopband 1) is very narrow (8.06-8.14 GHz), and the
higher (stopband 2) is wider (8.37-8.52 GHz). Between these two
stopbands  there is the band of backward waves. Stopband 2 is a
conventional lattice stopband which has ${\rm Re}\beta=0$.
Stopband 1 corresponds to the so-called {\it complex mode} which
is known in the theory for lattices of infinite wires (see e.g.
\cite{Engheta},\cite{Belov}). This complex mode is decaying and
the real part of the propagation factor is identically equal to
$\pi/d$. In our case, the relation ${\rm Re}\beta=\pi/d$ reflects
the fact that the directions of the dipole moments of the
scatterers are alternating along the propagation axis (i.e., two
adjacent isotropic scatterers have opposite polarizations).

Substituting relations \r{exact} and \r{DD} into Eq. \r{alp},  one
obtains the following explicit formula for the parameter $\alpha$,
\e \alpha (\omega ) ={\left[q_0-{2d_xd_y\eta\over \o}{\rm
Re}\left({1\over a_{mm}}\right)\right](\cos kd-\cos\beta d)+\sin
kd \over \sin\beta d}. \l{alpha} \f
Here $a_{mm} $ is a  function
of $\o$ for given geometrical parameters of scatterers and given
$\epsilon_b$. Since the frequency dependence of  $\beta(\o)$ is known from the
solution of the dispersion equation \r{disp1},  the frequency dependence of   $\alpha (\omega ) $
can be calculated from the above equation. In the next section we
will use this frequency dependence of   $\alpha (\omega ) $
 to find the explicit expressions for the effective material parameters
$\epsilon_{eff} (\omega ) $ and $\mu_{eff}(\omega ) $ of the homogenized lattice.

Our calculations show that $\alpha$ is also a crucial parameter
for the frequency bounds of the backward-wave range. The central
frequencies of both stopbands are frequencies at which $\alpha
(\omega )$ changes the sign. It is positive below 8.1 GHz,
negative in 8.1-8.45 GHz (backward-wave range) and positive again
above 8.45 GHz. Both stopbands are centered by the 2 zero-points
of $\alpha (\omega )$. Of course, both stopbands are within the
resonant band of each isotropic scatterer.

\section{Homogenization of the lattice of the isotropic scatterers }

Consider the frequency band $0<kd\le \pi$ within which
homogenization is possible and effective parameters
$\epsilon_{eff}$ and $\mu_{eff}$ can be, perhaps, introduced. To
determine the frequency dependence of the two parameters
$\epsilon_{eff} (\omega )$ and $\mu_{eff} (\omega )$, we needs to
find two relations (involving the propagation constant  $\beta
(\omega )$) between these two parameters. The first relation can
be obtained by  fitting the value $\beta/k$ with the effective
refraction index $n_{eff}=\sqrt{\epsilon_{eff}\mu_{eff}}$, i.e.,
one has \e \mu_{eff}={\beta^2\over k^2\epsilon_{eff}}.
\l{great2}\f Note that $k$ is proportional to $\o $ (i.e.,
$k=\o\sqrt{\epsilon_0\mu_0\epsilon_b}$) and  $\beta = \beta
(\omega )$ is given by the dispersion curve calculated in the
previous section.

To find another relation,  we
define the following averaged (over a cubic unit cell) electric and magnetic polarizations per unit volume:
$$
P={p\over V},\qquad
M={m\over V},
$$
where $V=d^3$ (in this section we assume that the lattice periods
are the same in all the three directions, i.e., $d_x=d_y = d_z =d$;
otherwise the lattice will be anisotropic). In the same way we can introduce
the following averaged electric and magnetic fields,
$$
<E>={1\over V}\int\limits_V E dV,\qquad <H>={1\over V}\int\limits_V H dV.
$$
Since the material considered here is isotropic, its wave impedance $Z$ can be expressed as  the
ratio $<E>/<H>$, i.e.,
 \e {<E>\over <H>}\equiv
Z . \l{taken}\f If we use the conventional expression
$Z=\sqrt{{\mu_0\mu_{eff}\over \epsilon_0\epsilon_{eff}}}$,  the
square root can be chosen either positive (in the range of forward
waves) or negative (in the range of backward waves; the ratio
${\mu_{eff}/\epsilon_{eff}}$ is always positive).  For a backward
wave in a continuous isotropic medium, the vectors $<\-E>=E\-x_0$
and $<\-H>=H \-y_0$ form a left-hand triad with the wave vector
$\-K=\beta \-z_0$ \cite{Veselago}, and thus the wave impedance
should be negative. Substituting \r{great2} into
$Z=\sqrt{{\mu_0\mu_{eff}\over \epsilon_0\epsilon_{eff}}}$, one
obtains \e Z=\sqrt{{\mu_0\over \epsilon_0}}{\beta\over
k\epsilon_{eff}} \l{zzz}.\f The above expression is more
convenient for use since it gives automatically a negative
impedance for a backward wave (when $\epsilon_{eff}<0$).

From the definitions for the effective material parameters, one
has
$$
\epsilon_0\epsilon_{eff}<E>=\epsilon_0\epsilon_{b}<E>+P,\qquad \mu_0\mu_{eff}<H>=\mu_0<H>+M .
$$
Using Eqs. \r{taken} and \r{zzz} and taking into account the
above relations, one can express the dimensionless parameter
$\alpha=\eta P/M$ (determined by Eq. \r{alpha}) as
$$ \alpha =\eta {P\over
M}=\eta{\epsilon_0(\epsilon_{eff}-\epsilon_{b})\over
\mu_0(\mu_{eff}-1)}{<E>\over <H>} =$$ \e
\eta{\epsilon_0(\epsilon_{eff}-\epsilon_{b})\over
\mu_0(\mu_{eff}-1)} Z    =  { (\epsilon_{eff}-\epsilon_{b})\over
\sqrt{\epsilon_b}  (\mu_{eff}-1)}  {\beta\over k\epsilon_{eff}} .
\l{great}\f The above equation gives the second required relation
between the two effective parameters.

Finally, from the two relations  \r{great2} and \r{great} we
obtain the following  expression for determining $\epsilon_{eff}
(\o) $   from the dispersion relation $\beta (\o) $, \e
\epsilon_{eff} (\o) ={{\beta^2\over k^2}+ {\beta\over
k\alpha}\sqrt{\epsilon_b}\over 1+ {\beta\over
k\alpha\sqrt{\epsilon_b}}}. \l{eps}\f Note that
$k=\o\sqrt{\epsilon_0\mu_0\epsilon_b}$ and  $ \alpha = \alpha
(\omega )$ is given by formula \r{alpha}. After $\epsilon_{eff}
(\o)$ is calculated, $\mu_{eff}(\o)$ is then calculated from
Eq. \r{great2}.

Fig. 6 shows the effective permittivity and permeability  for the
same lattice of isotropic scatterers as used before in Figs.
3$-$5. From this figure one sees that both $\epsilon_{eff}$ and
$\mu_{eff}$ are negative in the band of backward waves. This is
consistent to the Veselago theory.

In any of the two stopbands, the value $\beta$ contains non-zero imaginary
part and thus Eqs. \r{eps} and \r{great2} give complex values
for $\epsilon_{eff}$ and $\mu_{eff}$ at these frequencies. Since
the structure is lossless, $\epsilon_{eff}$ and $\mu_{eff}$ must
be real at the frequencies for which the homogenization is
possible. Therefore, we can conclude that the homogenization can
not be performed within the stopbands. That's why in Fig. 6 no
value is given for $\epsilon_{eff}$ or $\mu_{eff}$ in the
stopbands. From Fig. 6 one sees that outside the two stopbands the
absolute values of $\epsilon_{eff}$ and $\mu_{eff}$ are not very
high. The wavelength in the homogenized medium (which is equal to
$2\pi/\beta$) is not dramatically shortened within the passbands
(in which $\beta d<\pi$). This indicates that the present
homogenization model is self-consistent (unlike expressions
\r{MG1} and \r{MG2} derived from the Maxwell Garnett model). By
comparison Fig. 6 with Fig. 3, we can conclude that for a lattice
of resonant particles the Maxwell Garnett model  can not give
reliable values for either the real parts or the  imaginary parts
for the effective material parameters $(\mu_{eff})$ and
$(\epsilon_{eff})$. The Maxwell Garnett model can only predicts
qualitatively the location of the frequency band where the
effective permittivity and permeability have negative real parts.

\section{Conclusion}

In the present paper, we have presented an analytical model for an
isotropic negative meta-material, which is formed by a rectangular
lattice of isotropic scatterers with electric and magnetic
resonances. Each isotropic scatterer is formed by putting 3 pairs
of $\Omega$-particles on the opposite faces of a cubic unit cell
in an appropriate way. We have derived a self-consistent
dispersion equation and studied the dispersion properties of the
lattice.  The obtained dispersion curves are then used to
calculate correctly the effective permittivity and permeability in
the frequency band where the lattice can be homogenized. It is
interesting to see that the dispersion curves agree well with the
Veselago theory which predicts  backward waves when both
permittivity and permeability are negative. The Veselago theory is
non-complete (it was only for monochromatic waves and did not take
into account the dispersion of the medium) and some questions are
left open. It has been shown in the present paper that the
frequency range in which both the effective permittivity and
permeability are negative corresponds to the mini-band of negative
dispersion of the lattice at low frequencies within the resonant
band of the individual isotropic scatterer. It has been shown that
the frequency range of backward waves is determined by the  edges
of the two stopbands at low frequencies. We have compared the
results obtained by the present model with the previous results
obtained by  the Maxwell Garnett model, and shown that the latter
can not give reliable values for either the real part or the
imaginary part for any effective material parameter ($\mu_{eff}$
or $\epsilon_{eff}$) of the lattice. For lattices of resonant
scatterers,  the Maxwell Garnett model can only predict
approximately the location of the frequency band where the
effective permittivity and permeability have negative real parts.

The case when the lattice is formed by parallel bianisotropic
$\Omega$-particles has also been studied (in the appendix). It has
been shown that the bianisotropy of the lattice particles
suppresses those terms in the dispersion equation that can allow a
backward wave propagation and thus reduces the possibility for the
existence of backward waves.  Therefore, the present  formulas for
calculating correctly  $\epsilon_{eff}$ and $\mu_{eff}$ for the
negative meta-material from resonant isotropic scatterers (the
only isotropic negative meta-material that we know) is of importance.

\section*{Acknowledgment}

The partial support of  the Royal Swedish Academy of Sciences
is gratefully acknowledged. The authors are also grateful to Pavel A. Belov for some helpful discussion.

\setcounter{equation}{0} \renewcommand{\theequation}{A\arabic{equation}}
\section*{Appendix: The case of a  lattice of bianisotropic particles}

In this appendix,  we present briefly our studies for the case
when the lattice is formed by an infinite number of parallel
$\Omega$-particles.  We also explain why the magnetoelectric
coupling of these bianisotropic particles reduce the possibility
for the existence of backward waves.

We use a simplified wire-and-loop model for each
$\Omega$-particle, in which  we assume that the current flowing
around the loop is uniform (i.e., the electric polarization of the
loop is neglected).

If  the particle is excited by a $x$-directed  electric field
$E^{loc}$ (see Fig. 1), then the current in the loop and the
straight wire arms will be equal to the induced voltage in the
straight wire portion divided by the impedance for the series
connection of the loop impedance $Z_l$ and the straight wire
impedance $Z_w$, i.e.,
$$
I={{\cal E}\over Z_l+Z_w}={ E^{loc}l\over Z_l+Z_w}.
$$
Thus one obtains the following electric and magnetoelectric polarizabilities,
$$
a_{ee}\equiv{Il\over j\o E^{loc}}=-j{l^2\over Z_l+Z_w},
$$

$$
a_{me}\equiv{I\mu_0S\over E^{loc}}={\mu_0Sl\over Z_l+Z_w},
$$
where $S=\pi r^2$ is the loop area.

Now let the particle be excited by a $y$-directed local magnetic field $H^{loc}$. An electromotive force (i.e., voltage ${\cal E}$) is induced in the loop, and one has
$$
I={{\cal E}\over Z_l+Z_w}={ -j\o \mu_0SH^{loc}\over Z_l+Z_w}.
$$
Thus, one has electromagnetic and magnetic polarizabilities
$$
a_{em}\equiv{Il\over j\o H^{loc}}=-{\mu_0 lS\over Z_l+Z_w},
$$

$$
a_{mm}\equiv{I\mu_0S\over H^{loc}}={j\o \mu_0^2S^2\over Z_l+Z_w}.
$$

From the above expressions for the polarizabilities, the following
identities can be obtained: \e a_{ee}a_{mm}=a_{em}a_{me}, \l{id1}
\f \e a_{me}=-a_{em} . \l{id2}\f The second identity is due to the
reciprocity property of the particle. For such a bianisotropic
particle (local electric field is $x$-polarized and magnetic field
is $y-$polarized), we have (cf. Eqs. \r{p1} and \r{m1})
\begin{eqnarray}
p &=&a_{ee} E^{\rm loc}   +a_{em}  H^{\rm
loc} ,
\l{ppp} \\
m & =  & a_{me}  E^{\rm loc} + a_{mm} H^{\rm loc}.
\l{mmm}
\end{eqnarray}
In the rectangular lattice of parallel $\Omega$-particles, the
electric dipole moment of each particle is directed along the
$x$-axis and the magnetic dipole moment is directed along the $y$-axis
when the eigenmode with $\-E=E\-x_0$ and $\-H=H\-y_0$ propagates
along the $z-$axis (the local field has the same polarization as
the eigenmode). Substituting relations \r{ploc1} and \r{mloc1} into Eqs.
\r{ppp} and \r{mmm}, one obtains
\begin{eqnarray}
p(1-a_{ee}A-a_{em}D) &=& m(a_{ee}D +a_{em}{A\over \eta^2}),
\l{pp} \\
m(1-a_{mm}{A\over \eta^2}-a_{me}D) &=& p(a_{mm}D +a_{em}A). \l{mm}
\end{eqnarray}
The above two equations give
$$
(1-a_{ee}A-a_{em}D)(1-a_{mm}{A\over \eta^2}-a_{me}D)=(a_{ee}D +a_{em}{A\over \eta^2})(a_{mm}D +a_{em}A).
$$
Since all terms containing $D$ cancel out due to identities \r{id1} and
\r{id2}, one obtains
$$
1-a_{ee}A-a_{mm}{A\over \eta^2}=0,
$$
which can be rewritten in the following form \e {1\over
a_{ee}+{a_{mm}\over \eta^2}}=A. \l{bad}\f For a bianisotropic
particle, Eq. \r{sip} is generalized to the following
relation \cite{TV}, \e {\rm Im}\left({1\over a_{ee}+{a_{mm}\over
\eta^2}}\right)={\o^3 (\epsilon_0\mu_0)^{3\over 2}\over 6 \pi
\epsilon_0\epsilon_b}. \l{last}\f Substituting Eq. \r{exact} into
Eq. \r{bad} and using Eq. \r{last}, one obtains the following dispersion
relation, \e \cos \beta d=\cos k d+{\sin\beta d\over {X}},
\l{disp2}\f where
$$
X={2d_xd_y{\rm Re}\left({1\over a_{ee}+{a_{mm}\over \eta^2}}-q_0\right)\over \o\eta}.
$$
Eq. \r{disp2} has the same form as the classical dispersion
equation for a periodically loaded line \cite{Collin}), which is
also known as the Kronig-Pennie equation in the solid-state
physics. In the known cases described by this equation
(see e.g. \cite{Collin}) the negative dispersion exists only within the even
zones of frequencies ($2\pi n<kd<\pi(2n+1), \, n=1,2,3,...$). We
did not find  any physically realizable $\Omega$-particle  for
which the above dispersion relation would give a negative
dispersion within the first frequency zone ($0<kd<\pi$).

In the above analysis, the model for the $\Omega$-particle is
rough  (since the electric polarization of the loop is neglected)
and identity \r{id1} is approximate. We have studied the lattice
of parallel $\Omega$-particles using a more accurate model (as
used in \cite{JEWA}). This accurate model leads to another
dispersion equation (different from Eq. \r{disp2}) which is
quadratic with respect to $\cos\beta d$ and contains the
electromagnetic interaction factor $D$. Again, we did not find any
physically realizable  $\Omega$-particle  for which this
dispersion relation would give a negative dispersion within the
first frequency zone ($0<kd<\pi$). The error in Eq. \r{id1}
for real $\Omega-$particles is quite small and the terms
containing $D$ in this dispersion equation are relatively small.
Neglecting these small terms and removing the spurious root, we
obtained the same result as what Eq. \r{disp2} leads to.

Therefore, we can conclude that the bianisotropy of the lattice
particles    suppresses those terms in the dispersion equation
that can allow a backward wave propagation at low frequencies. The
possibility for the existence of backward waves is thus reduced
(it was reported with another reason in \cite{Marques} that the
bianisotropy is not good for the existence of backward waves).

\newpage

\section*{Figure captions}

\begin{description}
\item[Figure 1:]
\includegraphics[width=3.5in]{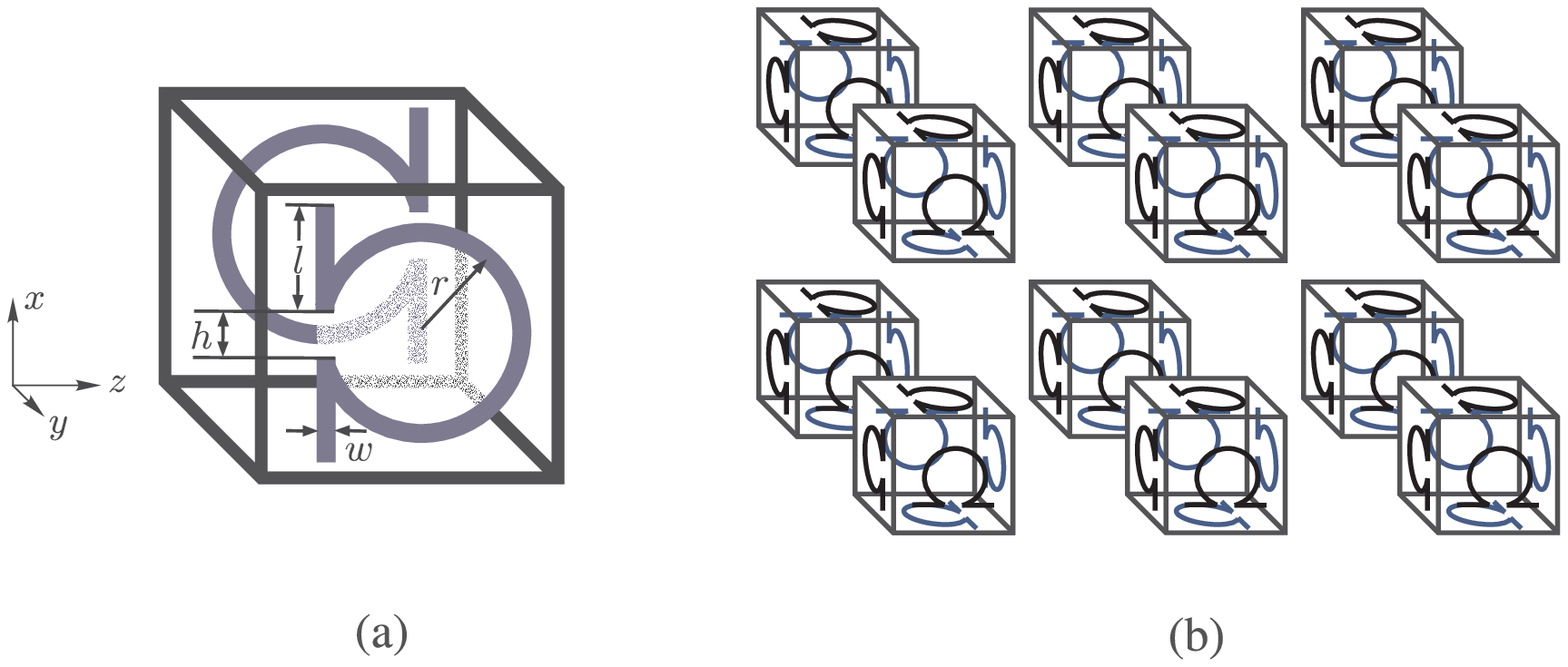}
(a)  Geometry for a cubic unit cell of $\Omega$-particles. For
graphic clearness, only 1 pair of $\Omega$-shaped perfectly
conducting particles are shown on 2 opposite faces of the cubic
unit cell. (b) A rectangular lattice of isotropic cubic unit cells
of $\Omega$-particles.

\item[Figure 2:]
\includegraphics[width=3.5in]{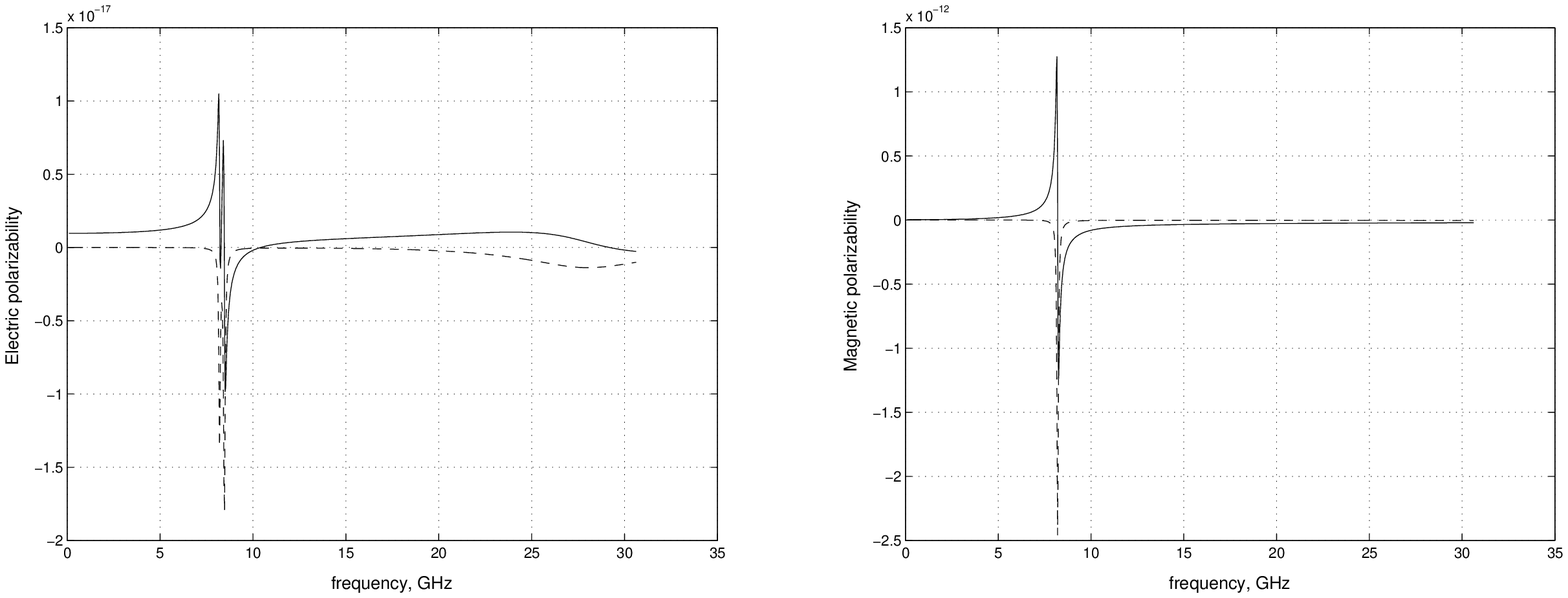}
Left: The real (solid lines) and imaginary (dashed lines) parts of the electric polarizability (left figure) and the magnetic polarizability (right figure) for an isotropic scatterer (a cubic
unit cell of $\Omega$-particles).

\item[Figure 3:]
\includegraphics[width=3.5in]{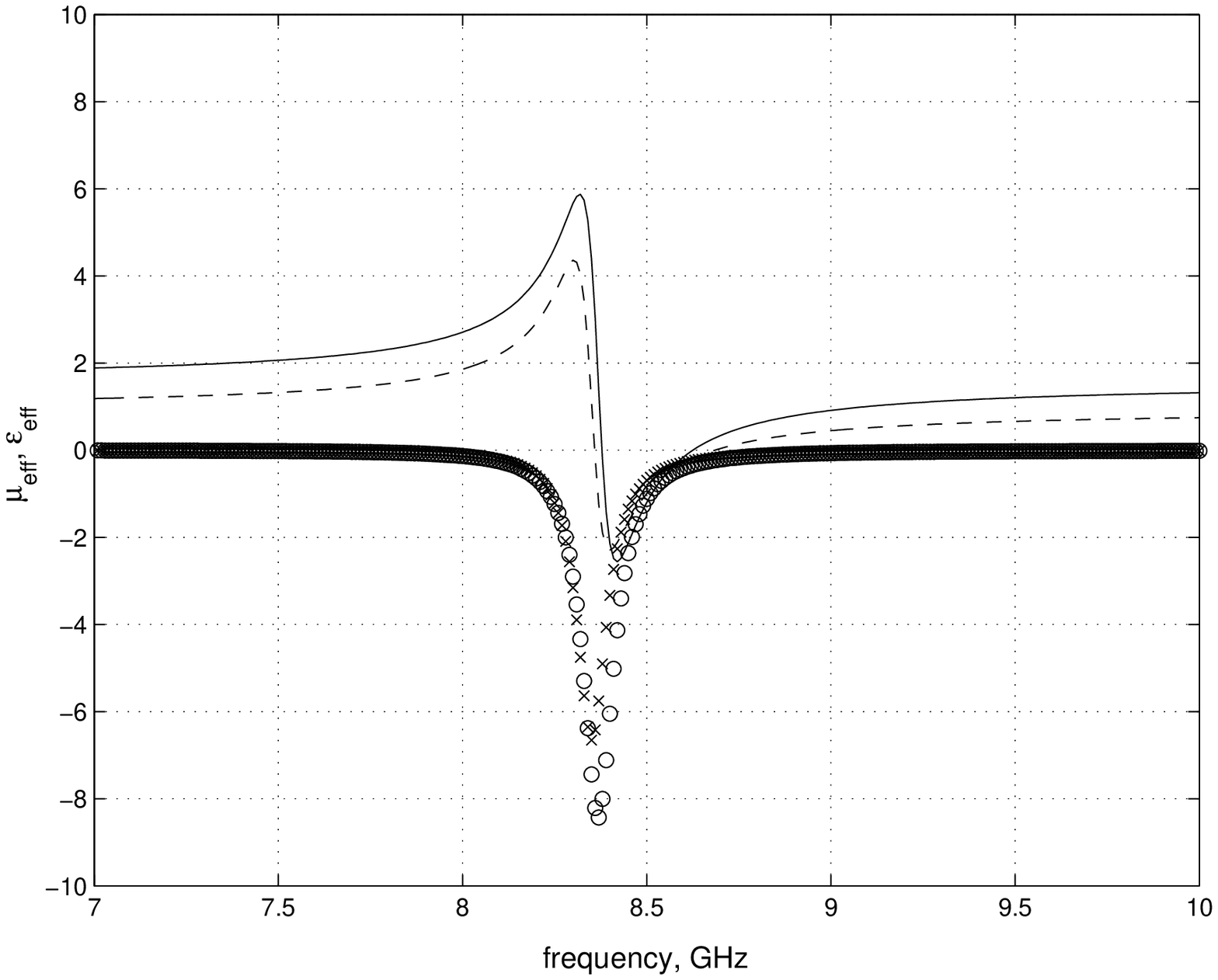}
The frequency dependencies of $\epsilon_{eff}$ and $\mu_{eff}$
predicted (incorrectly for the regular lattice) by the Maxwell
Garnett model  for  the negative meta-material formed by a lattice
of isotropic cubic unit cells of  $\Omega$-particles (as shown in
Fig. 1(b)). ${\rm Re}(\epsilon_{\rm eff})$ (solid line), ${\rm
Re}(\mu_{\rm eff})$ (dashed line), ${\rm Im}(\epsilon_{\rm eff})$
(circles), and ${\rm Im}(\mu_{\rm eff})$ (crosses) are shown.

\item[Figure 4:]
\includegraphics[width=3.5in]{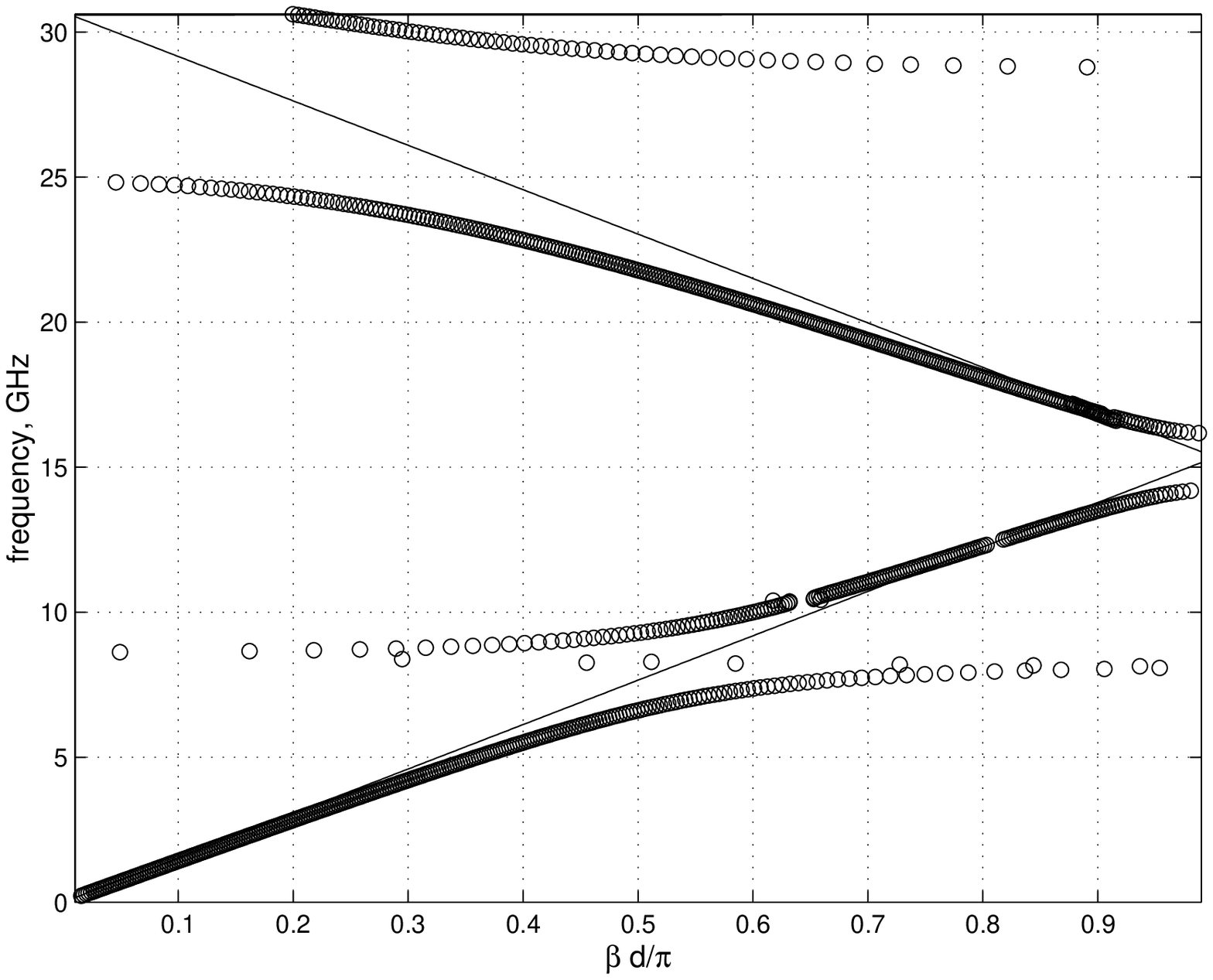}
The dispersion (indicated by the circles) for the same  lattice of
isotropic resonant scatterers as used for Fig. 3. The thin lines
indicates the dispersion of the homogenous background medium.

\item[Figure 5:]
\includegraphics[width=3.5in]{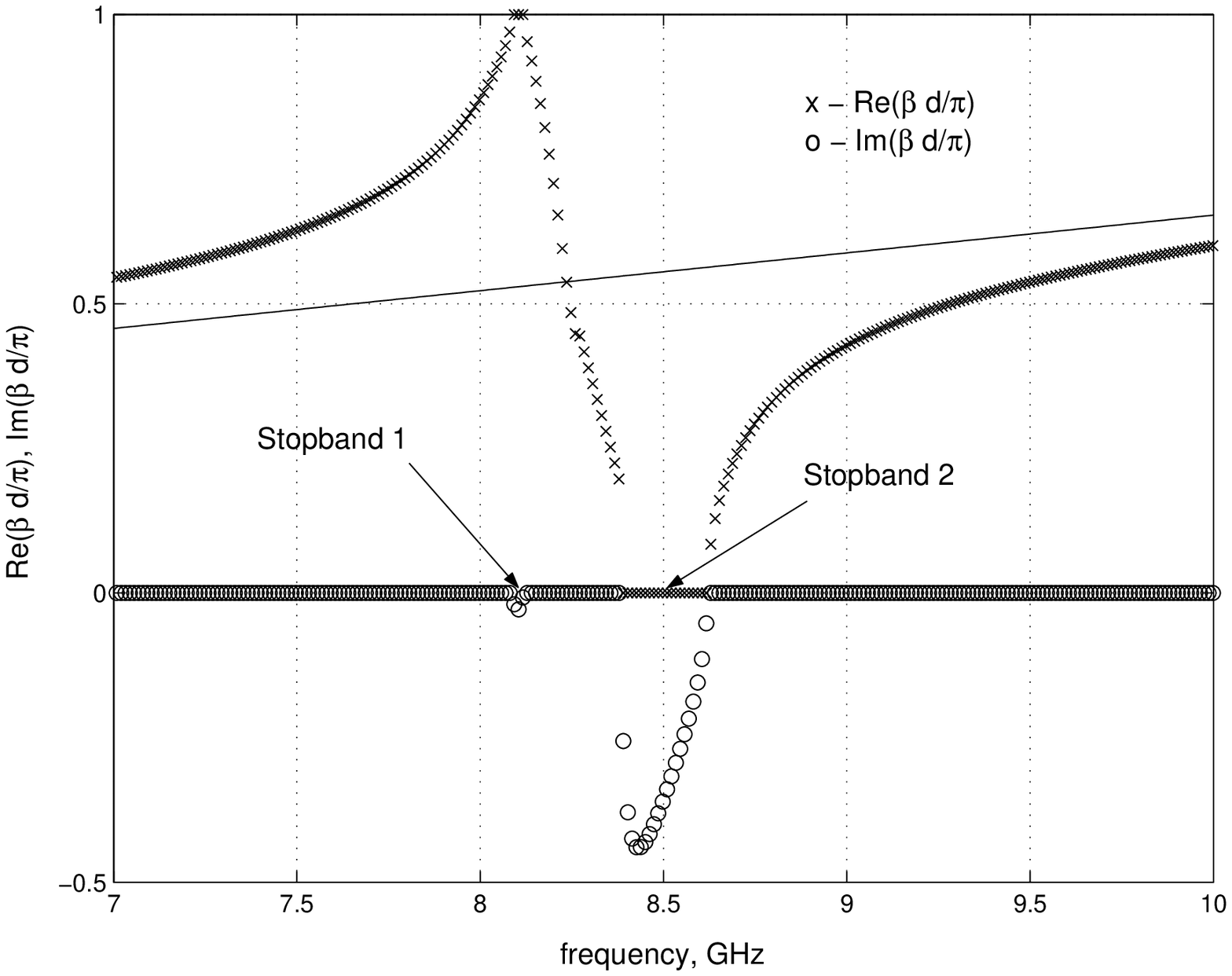}
An enlarged view of Fig. 4 for the frequency
dependence of ${\rm Re}\beta$ (crosses) and ${\rm Im}\beta$ (circles) near the resonance.
The thin line indicates the dispersion of the homogenous background medium.

\item[Figure 6:]
\includegraphics[width=3.5in]{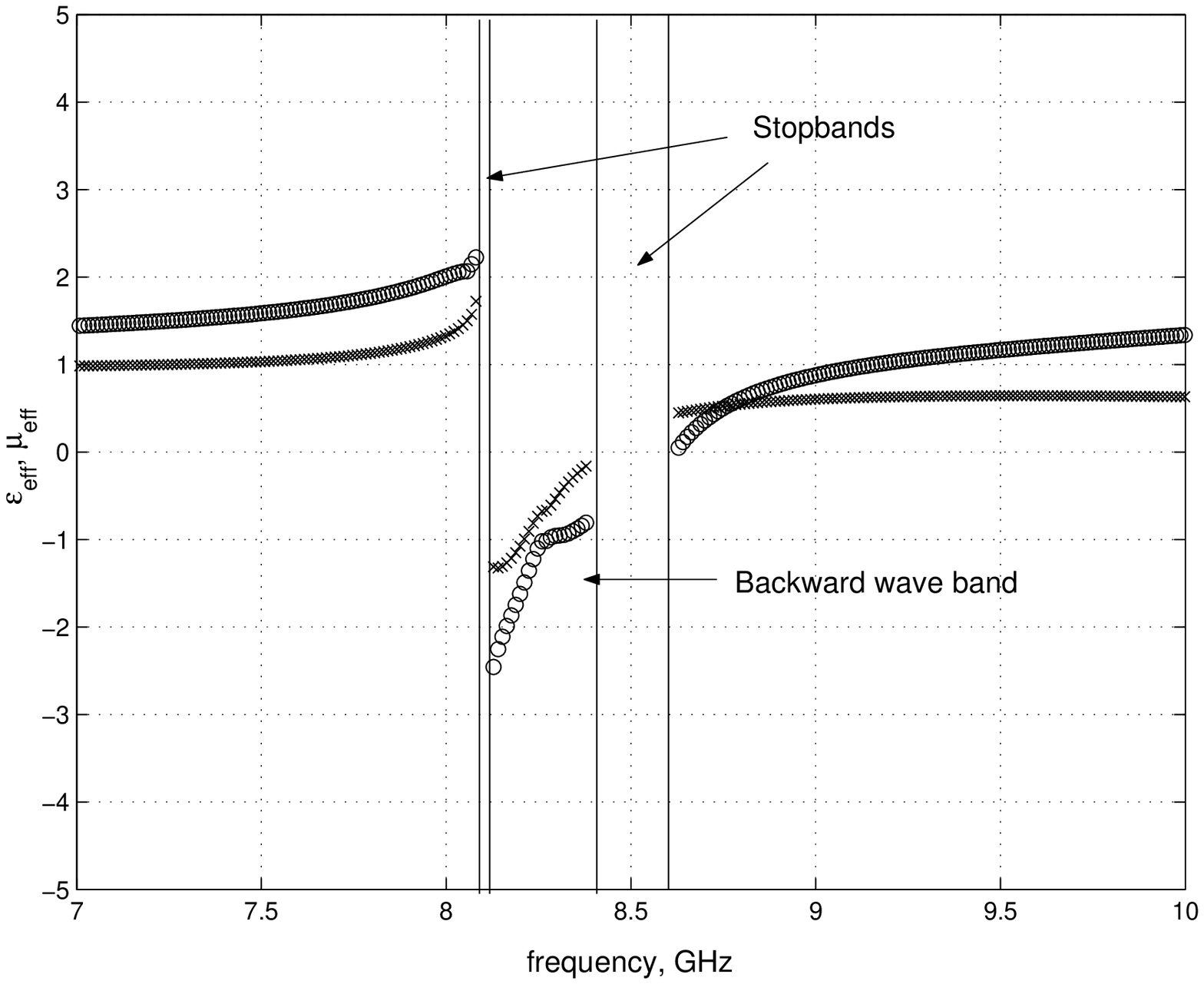}
The frequency dependencies of the effective permittivity
$\epsilon_{eff}$ (circles) and the effective permeability
$\mu_{eff}$ (crosses) for the same  lattice of isotropic resonant
scatterers as considered in Figs. 3$-$5.

\end{description}

\end{document}